\documentclass[letterpaper, 10 pt, conference]{ieeeconf}  % Comment this line out
\IEEEoverridecommandlockouts                              % This command is only
\usepackage{graphicx}
\usepackage{physics}
\usepackage{caption}
\usepackage{subcaption}
\usepackage{enumerate}
\usepackage{float}
\usepackage{dblfloatfix}
\usepackage{amssymb}
\usepackage{amsmath}

\usepackage[space]{cite}
\newtheorem{definition}{Definition}[section]
\title{\LARGE \bf Neural Network Embeddings for Test Case Prioritization}

\author{Lousada, João \\
Instituto Superior Técnico, Universidade de Lisboa \\
joao.b.lousada@tecnico.ulisboa.pt
\and
Ribeiro, Miguel \\
BNP Paribas \\
miguel.a.ribeiro@tecnico.ulisboa.pt % <-this % stops a space
}

\begin{document}

\maketitle
\thispagestyle{plain}
\pagestyle{plain}
\vspace{-1cm}

%%%%%%%%%%%%%%%%%%%%%%%%%%%%%%%%%%%%%%%%%%%%%%%%%%%%%%%%%%%%%%%%%%%%%%%%%%%%%%%%
\begin{abstract}

In modern software engineering, Continuous Integration (CI) has become an indispensable step towards systematically managing the life cycles of software development. Large companies struggle with keeping the pipeline updated and operational, in useful time, due to the large amount of changes and addition of features, that build on top of each other and have several developers, working on different platforms. Associated with such software changes, there is always a strong component of Testing. As teams and projects grow, exhaustive testing quickly becomes inhibitive, becoming adamant to select the most relevant test cases earlier, without compromising software quality.
We have developed a new tool called Neural Network Embeeding for Test Case Prioritization (NNE-TCP) is a novel Machine-Learning (ML) framework that analyses which files were modified when there was a test status transition and learns relationships between these files and tests by mapping them into multidimensional vectors and grouping them by similarity. When new changes are made, tests that are more likely to be linked to the files modified are prioritized, reducing the resources needed to find newly introduced faults. Furthermore, NNE-TCP enables entity visualization in low-dimensional space, allowing for other manners of grouping files and tests by similarity and to reduce redundancies.
By applying NNE-TCP, we show for the first time that the connection between modified files and tests is relevant and competitive relative to other traditional methods.
\vspace{0.25cm}

\end{abstract}

%%%%%%%%%%%%%%%%%%%%%%%%%%%%%%%%%%%%%%%%%%%%%%%%%%%%%%%%%%%%%%%%%%%%%%%%%%%%%%%%
\vspace{0.25cm}

\section{Introduction}

\textbf{Context} Given the complexity of modern software systems, it is increasingly crucial to maintain quality and reliability, in a time-saving and cost-effective manner, especially in large and fast-paced companies. This is why many industries adopt a \textit{Continuous Integration} (CI) strategy, a popular software development technique in which engineers frequently merge their latest code changes, through a \textit{commit}, into the mainline codebase, allowing them to easily and cost-effectively check that their code can successfully pass tests across various system environments \cite{santolucito2018statically}.

\textbf{Regression Testing} One of the tools used to manage software change is called regression testing. It is critical to ensure that the introduction of new features, or the fixing of known issues, is not only correct, but also does not obstruct existing functionalities. Regressions occur when a software bug causes an existing feature to stop functioning as expected after a given change and can have many origins (e.g. code not compiling, performance dropping, etc.), and, as more changes occur, the probability that one of them introduces a fault increases \cite{ShinThesis}. 

As software development teams grow, identifying and fixing regressions quickly becomes one of the most challenging, costly and time-consuming tasks in the software development life-cycle, rapidly inhibiting its adoption. Such teams often resort to modern large-scale test infrastructures, like core-grids or online servers \cite{Ziftci}. Consequently, in the last decades, there has been intensive research into solutions that optimize Regression Testing, accelerating fault detection rates either by alleviating the amount of computer resources needed or by reducing feedback time, i.e. the time delay between a software change and the information regarding whether it impacts the system's stability \cite{palma, Uber, yangJIT, learningfortcp, liang}.

The most prominent techniques for Regression Testing optimization are Test Case Minimization, where the number of tests is trimmed to avoid redundancies, Test Case Selection, where only a subset of all the tests is chosen, and Test Case Prioritization (TCP), where more relevant test cases are run first. This has become one thriving field, proven to have achieved meaningful results, with increasing research attention \cite{durelli, litrevtcp}.

More specifically, TCP aims to find the optimal permutation of test cases that matches a certain target, e.g. the ability to reveal faults as soon as possible, which is useful when there's a time budget or computer resources are limited \cite{ShinThesis}. The key goal of this study is to find out, \textit{a priori}, which test cases to prioritize, i.e. predicting which test cases will fail given a set of changes in the codebase. One possible solution would be to have a professional test engineer cherry-pick the most promising test cases. Unfortunately manual test case selection is time-consuming, counter-productive and error-prone, and is not scalable \cite{durelli}. Therefore, there has been a high demand for techniques that can automatically select test cases, minimizing human intervention \cite{Ziftci}. 

A good candidate strategy for TCP would be to first apply the tests that are directly related with the most recent commit. However, oftentimes, the relationship between the codebase and test case is not easily retrievable or is too complex to be inferred manually. Therefore, the focus of this work is to find relationships between the files modified in a given commit and the test cases more likely to suffer \textit{transitions}, i.e tests that were passing in previous commits and are now failing (i.e. \textit{regression}) and tests that were failing before and are now passing (i.e. \textit{progression}).

To achieve this, we propose a novel data-driven deep learning framework called NNE-TCP (Neural Networks Embeddings for Test Case Prioritization) that, based on historical data, is able to learn which modified files in a given commit led to test case transitions, and, from this information, deduce the aforementioned file-test case relationships. Thus, when new versions are created, our model should be able to predict, based on what files were modified in that commit, which test cases are more relevant to apply. In addition, NNE-TCP allows us to group files and test cases by similarity, thus allowing test case minimization.

%and, ultimately, rearranging tests in a smarter order.

In this paper our contributions are threefold: 

\begin{enumerate}
    \item To the best of our knowledge, this is the first time deep learning is applied to map file-test links for test case prioritization. We propose NNE-TCP, which uses Neural Networks Embeddings to select the most relevant test cases.
    
    \item We provide file and test case representation, based on similarity.
    
    \item We cross-check NNE-TCP's performance with other traditional methods of TCP.
\end{enumerate}

\textbf{Paper outline} Hereafter, Section 2 will correspond to the Background and Related Work providing all needed theoretical foundations. Then, Section 3 presents the framework of the NNE-TCP method. Section 4 provides the experimental validation of this new approach on a novel dataset, as well as a performance comparison to other traditional methods, threats to validity and future work. Section 5 summarizes and concludes the paper.

\section{Background and Related Work}

\subsection{Software Testing}

In software engineering, version control systems are a means of keeping track of incremental versions of files and documents, allowing the user to arbitrarily explore and recall the past commits that lead to that specific version\cite{santolucito2018statically}. Testing is a verification method used to assess the quality of a given software version. The building block of software testing is the test case, which specifies on which conditions the System Under Test (SUT) must be executed in order to detect a fault, i.e. for a given input, what are the expected outputs \cite{durelli}.

When test cases are applied, the outcome obtained is in the form of PASS/FAIL, with the purpose of verifying functionality or detecting errors. However, testing is very much like sticking pins into a doll - to cover its whole surface a lot of pins are needed, and the larger the doll, the more pins we require. Likewise, the larger and more complex the SUT, the greater the variety of test cases required. Therefore, to ensure that the health of the SUT is maintained throughout time, exhaustive testing is required to cover all possible scenarios \cite{7PrinciplesSoftTest}.

Inevitably, this task becomes impractical or even unfeasible due to the increasing complexity of the SUT, so testers have to find scalable approaches to counteract exhaustive testing, usually resorting to three techniques: Test Case Minimization, Test Case Selection and Test Case Prioritization (TCP), the latter being the target of this work. \\

\subsection{Test Case Prioritization}

As mentioned before, TCP rearranges test cases according to a given criteria, such as the probability of revealing faults.

\begin{definition}{\textbf{TCP}}
	Given the set of test cases, $T$, the set containing the permutations of $T$, $PT$, and a function from $PT$ to real numbers $f : PT \rightarrow \mathbb{R}$, find a subset $T'$ such that
	\begin{equation}
	    [f(T') \ge f(T'')], \hspace{0.5cm} \forall T'' \in PT.
	    %(T'' \neq T')
	\end{equation}
	In TCP, the function $f$ should be some relevant criteria such as code coverage, early fault detection, fewer resource demand, etc. \cite{ShinThesis}. 
\end{definition}

As an example, having five test cases (A-B-C-D-E), there are $5! = 120$ possible permutations, each with a different value in $f$. The goal of TCP would be to the one that maximizes that value.

%%%%% Option 1

One possible way of evaluating the optimal permutation is to compute the Average Percentage of Fault Detection (APFD) metric \cite{APFD}. 
\\

\begin{definition}{\textbf{APFD}}
	Let $T$ be the set of tests containing $n$ test cases and $F$ the set of $m$ faults revealed by $T$. Let $TF_i$ be the position, in a given permutation, of the first test case that reveals the $i^{th}$ fault \cite{ShinThesis}. Thus, APFD is defined as
\end{definition}

\begin{equation}
\text{APFD} = 1 - \frac{TF_{1}+\dots+TF_{n}}{nm} + \frac{1}{2n}.
\end{equation}

\par Simply put, higher values of APFD, imply higher fault detection rates, i.e. when APFD has value 1, all the failing tests are applied before all the ones that are passing, whilst when it has value 0, all the failing tests are applied at the end of the permutation.

In the literature, APFD is the most common metric for measuring TCP performance across different methods \cite{APFD}. However, for reasons that will become apparent later, our model detects transitions, not just regressions, so the APFD is unsuitable to assess the performance of NNE-TCP. For this reason, we define, for the first time, a metric for Test Case Prioritization based on test case transitions, the Average Percentage of Transition Detection (APTD).
\\
\begin{definition}{\textbf{APTD}}
	Let $T$ be the set of tests containing $n$ test cases and $\tau$ the set of $m$ transitions revealed by $T$. Let $T\tau_i$ be the order of the first test case that reveals the $i^{th}$ transition.
	
    \begin{equation}
        \text{APTD} = 1 - \frac{T\tau_{1}+\dots+T\tau_{n}}{nm} + \frac{1}{2n}.
    \end{equation}
\end{definition}

\vspace{0.5cm}

\par Therefore, similar to the APFD metric, if the APTD is $1$, all test cases that will suffer transitions are applied first, and if near $0$, all relevant test cases will be the last to be executed. By being able to create schedules that have a high APTD, we shorten the the time needed to both detect newly introduced regressions and find possible progressions.

%%%%%

\subsection{Machine Learning}

%The path to automatically solving a problem entails producing sets of instructions, that turn an input into a desired output, namely, algorithms. Nevertheless, Not all problems can be solved by traditional algorithms, 

Some problems can't be solved by traditional algorithms, due to limited or incomplete information. In our case, we don't know which tests are more likely to uncover faults. But this information is encoded within historical fault data, and it can be learned to enable future fault prediction. Hence, with the rise of data availability, there has been a growing interest in solutions that involve learning from data \cite{durelli}.

Benjamin Busjaeger \textit{et al.} \cite{learningfortcp} proposed an optimization mechanism for ranking test cases by their likelihood of revealing faults, in an industrial environment. The ranking model is fed with four features: code coverage, textual similarity, fault history and test age. These features were then used to train a Support Vector Machine model, resulting in a score function that sorts test cases. More recently, Palma \textit{et al.} \cite{palma} trained a Logistic Regression model, fed with similarity based features, such as similar lines of code. Liang \textit{et al.} \cite{liang} proved that prioritization at the commit-level, instead of test-level, would enhance fault detection rates, on fast-paced software development environments.

Another way of achieving effective TCP is using Semi-Supervised Learning approaches, like clustering algorithms. Shin Yoo \textit{et al.} \cite{Shinyoo} and Chen \textit{et al.} \cite{chen} endorse coverage-based techniques, claiming that making fast pair-wise comparisons between test cases and grouping them in clusters allows for humans to pick, more straight-forwardly, relevant non-redundant test cases, the assumption being that test cases that belong to the same cluster will have similar behaviours i.e. detect the same faults.

Deep Learning algorithms, in the context of TCP, have been gaining popularity in the last decade. Yang \textit{et al.} \cite{yangJIT}, following the work of Kamei \textit{et al.} \cite{kameiJIT}, implemented \textit{Deeper} to make \textit{Just-in-Time} prioritizations, powered by a Deep Belief Network framework used to generate more expressive features from the original dataset. Then, by feeding these features into a ML classifier, commits more likely to contain defects are predicted.

Recently, Spieker \textit{et al.} \cite{Spieker} were the first to implement a Reinforcement Learning approach to TCP, introducing Reinforced Test Case Selection (RETECS). This method prioritizes test cases based on their historical execution results and duration. RETECS has the advantage of being an adaptive method in a dynamic environment without compromising speed and efficiency.RETECS challenges other existing methods and has caught attention from other researchers, namely Wu \textit{et al.} \cite{time-window}.

\section{NNE-TCP}

In this section, we describe our novel approach, explaining how both file and test case embeddings can be learned from historical data, based on the assumption that files that cause similar tests to transition (i.e. to change states either through regression or progression), are similar to each other, and tests that fail together are likewise related. 
\par First, a brief description of embeddings is provided. Afterwards we present an in-depth walk-through of the implementation of NNE-TCP.

\subsection{Embeddings}
% Explains what embeddings are.
% Advantages in relation to other encoding
%techniques
% Why they are useful

The goal of embeddings is to map high-dimensional categorical variables into a low-dimensional learned representation that places similar entities closer together in the embedding space. This can be achieved by training a neural network.

\textit{One-Hot Encoding}, the process of mapping discrete variables to a vector of 0's and 1's, is commonly used to transform categorical variables, i.e. variables whose value represents a category (e.g. the variable \textit{color} can take the value \textit{red}, \textit{blue}, \textit{purple}, etc.), into inputs that ML models can understand. One-Hot encoding is a simple embedding where each category is mapped to a different vector (e.g. \textit{red}, \textit{blue}, \textit{purple} can correspond, respectively, to $[1,0,0]$ ,$[0,1,0]$ and $[0,0,1]$). 
\par This technique has two severe limitations: first, dealing with high-cardinality categories, (e.g. trying to map each possible color with this method would be unfeasible), and secondly, mappings are "blind", since vectors representing similar categories are not grouped by similarity (e.g. in this representation, the category \textit{purple} is no closer to \textit{blue} than it is to \textit{yellow}).
\par Thus, to drastically reduce the dimensionality of the input space and also have a more meaningful representation of categories, we could introduce \textit{embeddings}, lower dimensional vectors that represent categories by mapping similar categories to similar vectors. For example, we could map the variable \textit{color} to a lower dimensional space, \textit{red} $ =[1,0]$, \textit{blue} $ =[0,1]$ and \textit{purple} $ =[1,1]$, by taking advantage of the fact that \textit{purple} is a combination of \textit{red} and \textit{blue}.

% Explains how Embeddings are learned
In our case, we take each file and each test and represent them as n-dimensional vectors, with the goal of representing similar files and similar tests as similar vectors. The key aspect of embeddings is that these n-dimensional vectors are trainable, which means that each vector component can be adjusted in order to push vectors representing of related objects together. As a result, after training, the supervised learning task will be able to predict whether two categories are similar.

%\par To achieve this, during training, the embedding representation will gradually improve, with gradient descent, by minimizing a cost function. As a result, it will be able to produce more accurate predictions of file-test links and more meaningful entity representations, squeezing together entities that have similar behaviour, i.e. test cases whose transitions always occur simultaneously, will have corresponding embedding vectors pointing in the same direction \cite{geron_hands_on}. 

%In the case of a neural network embedding, its parameters, the weights, are the embedding components, which are adjusted by backpropagation. 

\subsection{Implementation}\label{imp}
% Approach Steps
% Model Description

Our NNE-TCP approach is sustained by a predictive model that tries to learn whether a modified file and a test case are linked or not. After training, the model can be used to make new predictions on unseen data and create test schedules that prioritize test cases more likely to be related with files modified in a given commit. The implementation of this algorithm was done with the Python Deep Learning Library - \textit{Keras} \cite{chollet2015keras} and was adapted from \cite{osinga2018deep}. The steps taken to develop the framework were: 

\begin{enumerate}
    \item Load and Clean Dataset.
    \item Create Training Set.
    \item Build Neural Network Embedding Model.
    \item Train Neural Network Model.
    \item Evaluate Model's Performance.
    \item Visualize Embeddings using dimensionality reduction techniques. 
\end{enumerate}

% 1)
\vspace{0.4cm}
\textbf{1)} The dataset should contain records of the files modified in every commit, as well as test cases that suffered transitions.  
Data cleaning is increasingly important as development environments become more dynamic, fast-paced and complex. Dealing with modified file and test case records will inevitably have noise. For example, in a dynamic software development environment, files can become outdated, deprecated, duplicated or renamed.
\par For this reason, eliminating redundant files and tests as well as removing files and tests that haven't been modified or transitioned recently are necessary steps to obtain a cleaner dataset.
It's worth mentioning that another source of noise in the data arises from the fact that some files and tests will be linked by chance and will not correspond to actual connections. For this reason, we need to further remove these connections, a step that will be described later.

% 2)
\textbf{2)} After loading and cleaning the dataset, in order for the model to learn the supervised learning task, it needs to be trained. Like any other Machine Learning problem, the dataset must be split into two sets: training, for the algorithm to learn from known examples, and testing, used to obtain an unbiased evaluation of the trained model. The size of each set can vary, depending on the dataset size and characteristics. In this work, we used a ratio of $80/20 \%$ between the training and testing sets, in order to maximize the amount of data available for training, without compromising an unbiased evaluation.

The problem we are trying to solve can be stated as: given a file and a test case, predict whether the pair is linked, i.e. predict if the modification of a given file could impact the outcome of a given test case's execution, based on the commit history. Subsequently, the training set will be composed of pairs of the form: $(file, test, label)$. The label will indicate the ground-truth of whether the pair is or is not present in the data. In any of the commits, if there is a test case that suffered a transition on a commit where a given file was modified, then that file and test case constitute a positive pair.

To create the training set, we need to iterate through each commit and store all pair-wise combinations of files and test cases. However, there will be commits in which many files were modified, or where many test cases suffered transitions, which means that the training set will contain many false-positive pairs, i.e. positive pairs that do not represent an actual file-test link. This aspect can be mitigated by removing from the training set pairs that only occur a very small number of times, while preserving those that reoccur repeatedly throughout the commit history.

Because the dataset only contains positive examples of linked pairs, in order to create a more balanced dataset we need to generate negative examples, i.e. file-test pairs that are not linked. A negative example constitutes test case that is not linked to a modified file. This occurs when a certain file is modified, but a test case status remained unchanged, i.e. it did not transition. It's important to emphasize that, in a given commit, the files that were not modified in that commit and the test cases that transitioned do not constitute a negative pair, since there is no guarantee that if that file were to be modified, it would not impact the test case in question. For this reason, we only consider negative pairs of the form: (file modified, unaltered test case). Thus, to keep a balanced dataset, for any given commit with a certain number of positive pairs, we randomly pick a proportional number of unique negative pairs.

In our dataset, for the positive samples, the $label$ is set to 1, whereas for negative examples, the $label$ is set to $0$ when the task at hand is classification or $-1$ when we're using regression, for reasons that will become apparent later.

% Data Generator 
As a result of having to create balanced examples for every commit and specially when dealing with large datasets, it could become unpractical, in terms of memory and processing power, to generate and store the entire training set at once. Consequently, to alleviate this issue, the \textit{Keras} class \textit{Data Generator} offers an alternative to the traditional method, by generating data \textit{on-the-fly} - one batch (i.e. subset of the entire dataset) of data at a time \cite{chollet2015keras}.

%new paragraph
In our case, each batch corresponds to the pairs that resulted from a single or the association of several commits, where a variable amount of files and test cases are involved. Then, iteratively, each batch is fed to the model, so that the weights can be adjusted to minimize the cost function. Once the weights are updated for every batch, we say that an epoch has passed i.e. after training each batch sequentially, the entire dataset is covered. The Data Generator class has several parameters that need to be adjusted: the batch size, comprising the number of commits in a single batch; the shuffle flag, that, if toggled, shuffles the order of the batches once a new epoch begins; and the negative-ratio, which represents the dataset class balance ratio e.g. if set to 2, there will be twice as many randomly picked negative examples as there are positive.

% 3)
\textbf{3)} Having created the training set, the next step is to build the learning model's architecture. The inputs of the neural network are (file, test) pairs and the output will be a prediction of whether or not there is a link. The \textit{Keras} Deep Learning model is depicted below in Fig. \ref{keras_model} and is composed of the following layers: \\

\begin{itemize}
    \item \textbf{Input}: 2 neurons. One for each file and test case in a pair.
    \item \textbf{Embedding}: map each file and test case to a n-dimensional vector.
    \item \textbf{Dot}: calculate the dot product between the two vectors, merging the embedding layers.
    \item \textbf{Reshape}: reshape the dot product into a single number.
    \item $\lbrack$Optional$\rbrack$ \textbf{Dense}: generate output for classification with sigmoid activation function. Commonly known as the logistic function $S(x) = 1/e^{-x} +1$.
\end{itemize}

\begin{figure}[h]
\centering
\includegraphics[width=\columnwidth]{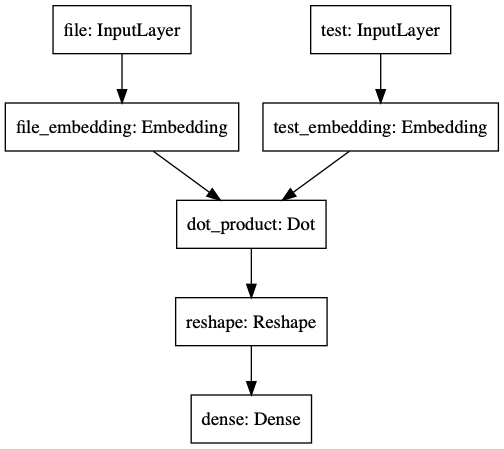}
\caption{Neural Network Embedding Model Architecture}
\label{keras_model}
\end{figure}

\vspace{0.4cm}

The file and test case in a pair are fed as inputs to the neural network. Afterwards, the embedding layer deals with representing each one of them as a vector in n-dimensional space. Then, we combine the two representations into a single number by calculating the dot product between the file and test embedding vectors.

\par For example, suppose file A and test Z are linked, i.e. the pair (A, Z) has label $1$. If file A is represented by the embedding vector $[1,3,1]$ and test Z by $[-0.5,0,1]$, then the dot product between the two is $0.5$. 
This number will be the output of the model, for a regression task, which will then be compared with the true label of the pair (in this example, $1$). The weights (i.e. the components of the embedding vectors) will then be adjusted accordingly, in a way as to increase the dot product to match the label observed.

In order to train the model, we need a metric to measure how close the model's output is to the label, called the loss function.

For the regression task, we use the mean squared error (MSE), defined as
\begin{equation}
\text{MSE} = \frac{1}{n}\sum_{i=1}^{n}(y_i - \hat{y}_i)^2,
\end{equation}
where $n$ is the number of predicted samples and where $y$ and $\hat{y}$ correspond to the true and predicted labels, respectively. Our goal is to minimize this loss function by readjusting the embedding vectors' values such that the dot product between the file and test embedding vectors, $\hat{y}$, becomes closer to the correct label, $y$.

For classification, because the label is either $0$ or $1$, a Dense layer with a sigmoid activation function needs to be added to the model to squeeze the output between $0$ and $1$. The chosen loss function was the binary cross-entropy (BCE), defined by
\begin{equation}
    \text{BCE} =-\frac{1}{n}\sum_{i=1}^{n}\lbrack y_i\log(\hat{y}_i) + (1-y_i)\log(1-\hat{y}_i)\rbrack,
\end{equation}
which measures the similarity between binary classes \cite{chollet2015keras}. \\

\textbf{4)} Having built the model's architecture, the next step is to train it with examples produced by the Data Generator, for a certain number of epochs. At this stage, the weights are updated such that the respective loss function (depending on whether we are using classification or regression) is minimized and the accuracy when predicting whether the pair is positive or negative is maximized. If the algorithm converges, the model is ready to make predictions on unseen data and produce meaningful representations of file and test case embeddings.

% 5)
\textbf{5)} After training the model, we are able to make predictions on new, unseen commits. We can validate the true accuracy of our model using the test set, which corresponds to $20\%$ of the entire dataset. To evaluate the model's performance, we measure its APTD as well another metric, which varies depending on the supervised task at hand, and which we define next.

\par For regression tasks, the metric used is the Mean Absolute Error (MAE) that calculates the average of the difference between the true and predicted labels and it is defined as

\begin{equation}
\text{MAE} = \frac{1}{n}\sum_{i=1}^{n}\abs{y_i - \hat{y}_i}.
\end{equation}

For classification tasks, we instead use the Binary Accuracy, that counts the frequency of correctly predicted labels, i.e. how often $y_i$ matches $\hat{y}_i$, and it is defined as,

\begin{equation}
\text{Accuracy} = \frac{TP + TN}{TP + FP + TN + FN},
\end{equation}
where $TP$ and $TN$ mean true positive and true negative and occur whenever $y_i = \hat{y}_i$, either for positive or negative labels, whereas $FP$ and $FN$ occur when $y_i != \hat{y}_i$.

When evaluating the model with the test set, given that it does not know which tests will suffer transitions, the files modified on each commit will have to be paired with every test.  
For example, if three files were modified and we have $4,000$ test cases, there will be $3 \times 4,000 = 12,000$ pairs in that commit.
Then the algorithm will rank all test cases by the likelihood of them being linked to each modified file, a value that is obtained from the dot product between each of the modified files' embedding vectors and each test's. This will result in a matrix of scores $m \times n$ where $m$ is the total number of test cases and $n$ corresponds to the number of files that were modified in the current commit.
\par Subsequently, to create a test ordering, we pick the maximum value from each row, resulting in a single vector of size $m$. Test cases are then ranked by descending order of their score. Following the example above, $m=4,000$ and $n=3$ and the result will be a $n$-dimensional vector of prioritized test cases.

From this ordering of tests, we can calculate the APTD metric by applying test cases according to their rank in the prioritization generated.

% 6)
\textbf{6)} Lastly, a useful application of training embeddings is the possibility of representing the embedding vectors in a reduced dimensional space, providing a helpful intuition about entity representation. 
\par Since embeddings are represented in an n-dimensional manifold, one has to resort to manifold reduction techniques to represent elements in 2D or 3D, in order for the manifold to be understandable by humans. This can be done, for example, by using Uniform Manifold Approximation and Projection (UMAP) \cite{umap}, a technique used to map high-dimensional vectors to lower dimensional spaces, while preserving the structure of the manifold and the relative distances between elements.

%\begin{figure*}[t]
%\centering
%\hspace*{-0.5cm}  
%\includegraphics[width=\textwidth]{figures/3d_data_clean_tests.pdf}
%\caption{Data Cleaning shows Average number of occurrences per files/tests, i.e. Number of Modifications per File (MpF) and Number of Transitions per Test (TpT) as a function of Date, Individual Threshold and Threshold Pairs.}
%\label{surf}
%\end{figure*}
\section{Experimental Setup}

An experimental evaluation of our approach to the algorithm is presented in this section, with a brief description of the dataset used, the corresponding cleaning steps applied and the results obtained. %and comparison with traditional methods.

\subsection{Data Description} 

The dataset used to train NNE-TCP consists of historical data collected from a versioning software log, i.e. document that displays commit log messages, belonging to a large company of the financial sector. The data was collected over a period of four years and contains information relative to the commits, tests and modified files. The dataset contains over 4000 commits, 10800 files and 4000 test cases.

\subsection{Data Cleaning}

For the data cleaning process, we took three aspects into account:
\begin{itemize}
    \item \textbf{Date Threshold}: timestamp from which we consider file/test modifications/executions - if a file/test has not been modified/executed for $n$ months, it is considered deprecated and is removed.
    \\Values considered: $n=[6, 12, 18, 24, 30, 36]$ expressed in months.
    \item \textbf{Individual Threshold}: the individual frequency of each element - files and tests that appear fewer than $n$ times are removed, because they are likely to be irrelevant.\\Values considered: $n=[0, 2, 4, 6, 8, 10]$.
    \item \textbf{Pairs Threshold}: frequency of file/test pairs - pairs that occur fewer than $n$ times are likely to have happened by chance, so they are removed.\\Values considered: $n=[0, 1, 2, 5]$
\end{itemize}

To remove the noise from the data we need to remove files, tests and pairs that rarely are part of a commit in the dataset. A file/test is said to occur in a commit, if it was modified/executed. We want the average number of occurrences per file/test to be larger, leading to a higher density of relevant files and tests, in order to obtain a higher quality dataset.

%The density varies according to the surface plot depicted in Fig. \ref{surf}.In the first plot, we can see that the parameter that most influences the average modifications per file is the individual threshold, that, for higher values, only preserves the files modified most often. However, from the initial 10800 files, only 300 remain. So there is a significant compromise between having an expressive dataset with only very frequent pairs and having a broader scope of valid commits. In both plots, for a pair threshold of 1, it is possible to see increases in the density, whilst making sure relevant files and tests are not being dropped. \par Given the description above, in order to have a more expressive dataset without significantly reducing its size, the data cleaning values chosen are: 

After exploring the possible values described above, we found that the combination of parameters that minimized noise without compromising the generalization of the dataset were:

\begin{itemize}
    \item \textbf{Date Threshold}: 12 months.
    \item \textbf{Individual Threshold}: 5 occurrences.
    \item \textbf{Pairs Threshold}: 1 occurrence.
\end{itemize}

\subsection{Traditional Methods}
To validate the efficiency of NNE-TCP method, we compare it to three test case prioritization methods: 
\begin{itemize} 
    
    \item \textbf{Random} - each test is assigned a random prioritization. This method will serve as a baseline for comparison.
    \item \textbf{Transition} - a fixed prioritization of the test cases is used across every commit. Tests are ranked by their rate of transition, determined by the number of times they have suffered transitions in the past.
    \item \textbf{History} - tests that suffered transitions more recently are assigned a higher rank. 
    
    %i.e. we rank the test cases based on the parameter $N_i$, which measures the number of commits past since the test case $t_i$ suffered a transition (it is set to $0$ whenever the test transitions again). Thus, the smaller the value of $N_i$, the more recently the test transitioned, and therefore, the higher its rank.
    This approach is based on two assumptions: regressions are likely to be fixed quickly, i.e. a test that just started failing should transition through a progression soon, and stable test cases, i.e. tests that rarely have any transitions are less relevant. Furthermore, while a project is in state of active development, it is likely that only the same subset of test cases will be involved in transitions, which will be prioritized by \textit{History}. However, this approach does not take into account the fact that a test case that has been failing for some time is more likely to be fixed than a test that had just started failing.
\end{itemize}

\section{Results}

\subsection{Training and Fine-Tuning}

After framing the problem and having the data cleaned, we can move on to training the model. This training consists of learning the embedding representation, by updating the neural network's weights, through backpropagation. This process is then repeated for 10 epochs, collecting the respective metrics, Accuracy or MAE, depending on the supervised learning task at hand. The algorithm is then evaluated by measuring the APTD for test orderings on the test set. 
\par In this training process, we have some hyper-parameters that influence the model's performance, and these need to be fine-tuned to reach the best results. Some examples of hyper parameters are the embedding size, batch size and negative-ratio. Fine-tuning analysis allows us to find the best combination of parameters that maximize our metrics, by experimenting different values. To determine them, a grid search was conducted, covering different combinations between possible values for each parameter. The choice of optimizer function was not subject to fine-tuning but it is indicated that Stochastic Gradient Descent (SGD) was used.
\par The parameters that were found to produce the best results are summarized in Table \ref{params}. Due to limited time and computer power, the values obtained are most likely sub-optimal and more refinement is needed to reach optimality.
%not the optimal values, given the degree of refinement, but rather sub-optimal.
Hereafter, unless stated otherwise, the same set of parameters is used throughout the paper.

\begin{table}[h]
\small
\centering
\begin{tabular}{lcc}
\hline
\multicolumn{1}{c}{\textbf{Parameter}} & \multicolumn{1}{l}{\textbf{Best Value}} & \textbf{\begin{tabular}[c]{@{}c@{}}Possible Values\end{tabular}} \\ \hline
Embedding Size                         & 200                                     & {[}50, 100, 200{]}                                                  \\
Negative Ratio                         & 1                                       & {[}1, 2, 3, 4, 5{]}                                                 \\
Batch Size                             & 1                                       & {[}1, 5, 10{]}                                                      \\
Epochs                                 & 10                                      & {[}10, 100{]}                                                       \\
Task                                   & Regression                              & {[}Classification, Regression{]}                                    \\
Optimizer                              & SGD                                     & SGD                                                                 \\
Date (months)                          & 12                                      & {[}6,12,18,24,30,36{]}                                              \\
Individual                             & 5                                       & {[}0,1,2,3,5,10{]}                                                  \\
Pairs                                  & 1                                       & {[}0,1,2,5{]}                                                       \\ \hline
\end{tabular}
\caption{Best combination of parameters obtained after grid-search fine tune analysis and possible values used.}
\label{params}
\end{table}

\begin{figure}[H]
\centering
\hspace*{-0.5cm}  
\includegraphics[width=\columnwidth]{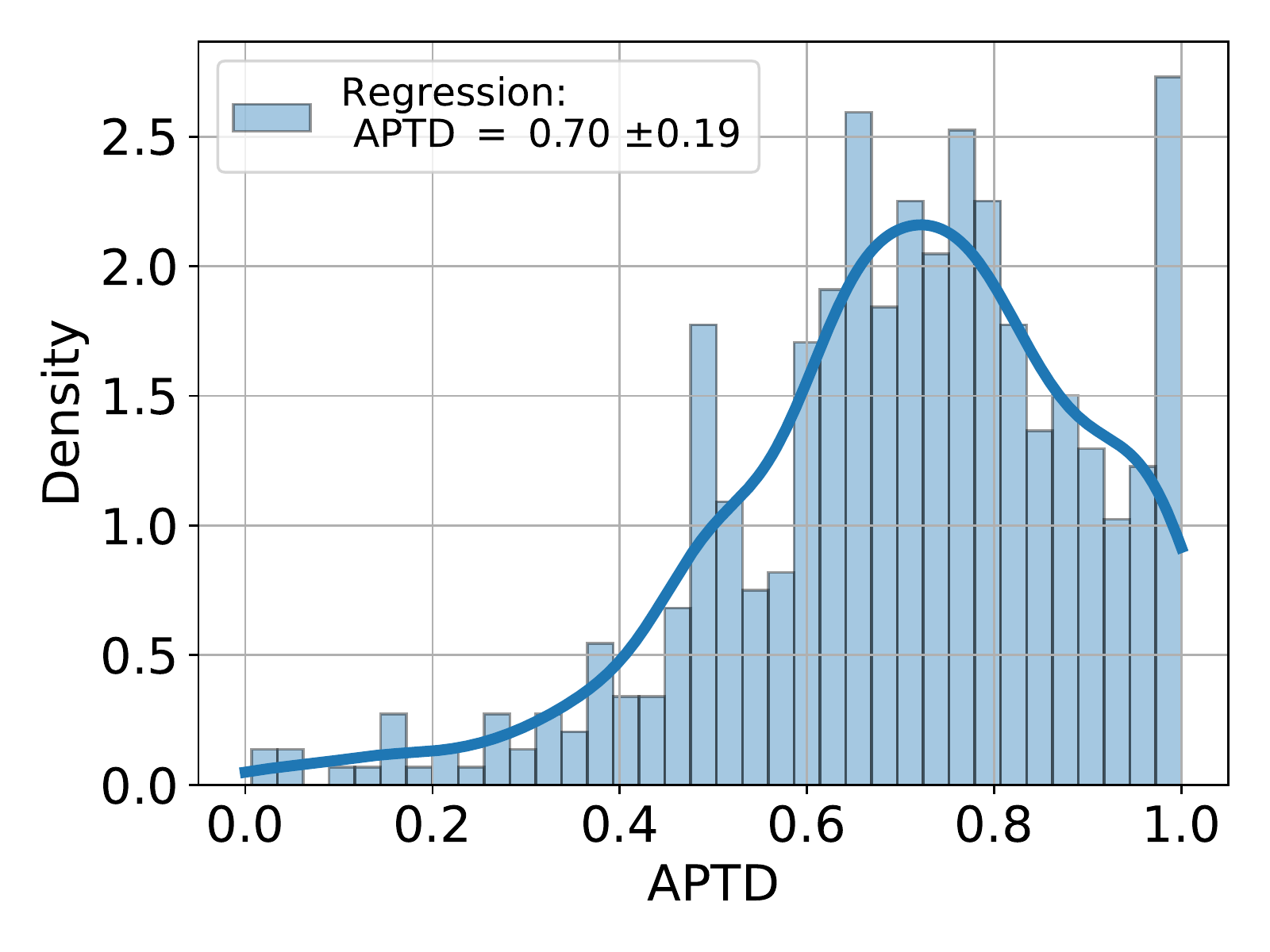}
\caption{APTD histogram and density distribution function (Blue line) for optimal parameter combination}
\label{best_result}
\end{figure}

By training the model with the aforementioned combination of parameters, we were able to obtain an APTD $= 0.70 \pm 0.19$. These results are represented in Fig. \ref{best_result}, in which we represent a histogram with a Gaussian Kernel distribution - a continuous probability density curve - of the different APTD values obtained for the different commits of the test set.

From Fig. \ref{best_result} we can conclude that the model managed to produce a desirable result, with a fairly high APTD. It can also be seen that highest bar in the histogram corresponds to an APTD near 1, which may indicate that, for some commits, the algorithm was able to correctly predict which tests would suffer transitions and prioritized them first.

\subsection{Cross-Validation}

Having fine-tuned the model's parameters to maximize the its performance, the next step is to account for over- and underfitting. The former occurs whenever a model is not able to generalize from the particular set of data, becoming biased. The latter indicates that the model is unable to capture the underlying structure in the data. Our goal is to make sure that, during training, the model can generalize known examples, to predict new information it has never seen before.

%while ensuring that the model is not complex enough to adjust to the nuances present in the data.
\begin{figure}[t]
\centering
\includegraphics[width=\columnwidth]{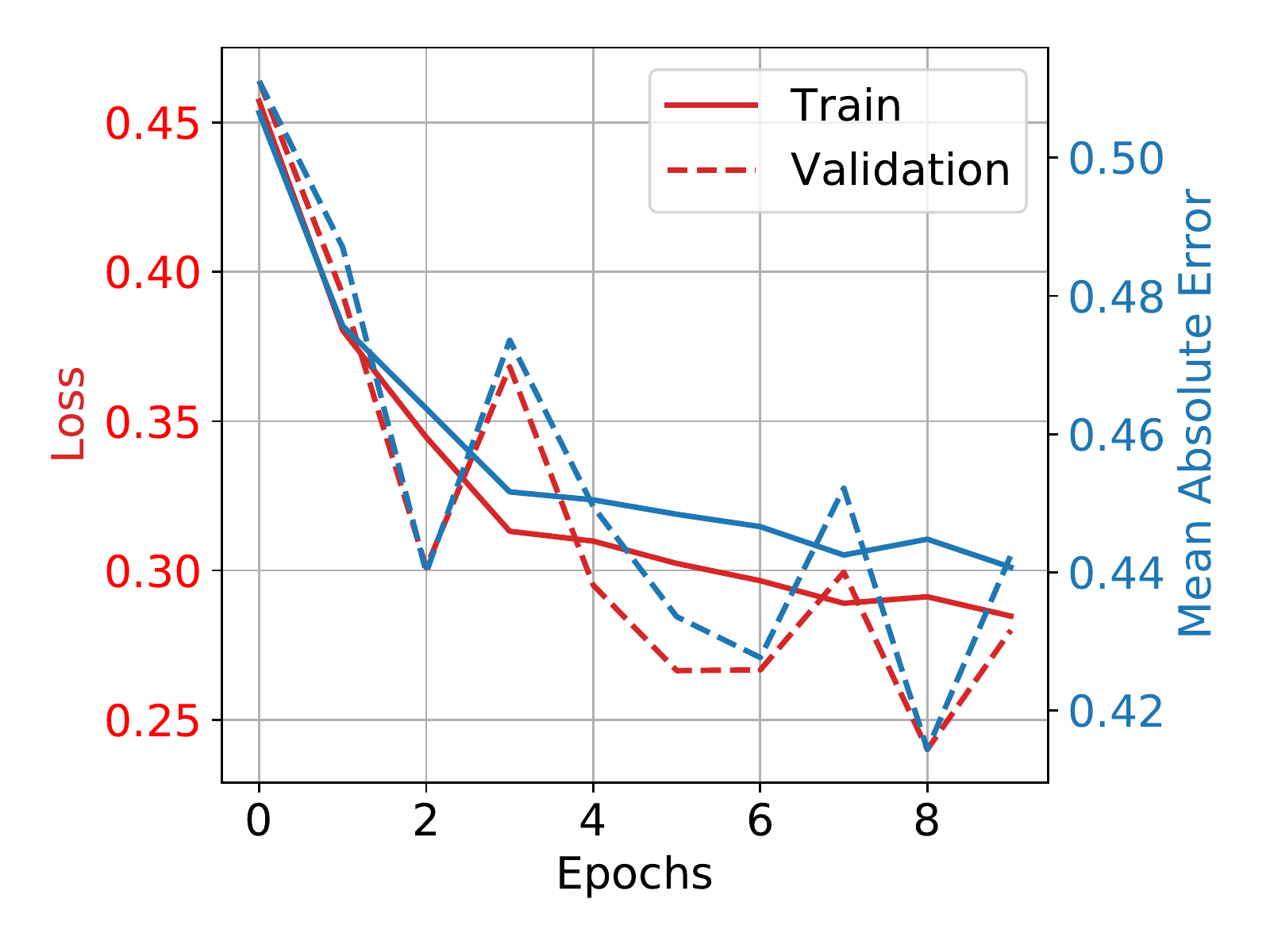}
\caption{Loss (red lines) and MAE (Blue lines) for Cross Validation. Full and Dashed lines correspond, respectively, to the training and validation set. Dashed lines should not deviate from full lines.}
\label{cv}
\end{figure}

Using \textit{Scikit Learn}'s K-Fold cross validation feature \cite{scikit-learn}, the training set can be divided into 10 separated subsets (the gold standard value), called \textit{folds}, used to reduce the bias of a model's fit on the training set. The model is trained ten times, and while nine of those folds are used for training, the one fold is used for validation. Afterwards, we average the ten obtained results for each fold, creating a single model ready to make predictions on unseen data and, hopefully, with enough generalization capability.
The one fold is called the \textit{validation-set} and, if the model is not over- or underfitting the data, the loss function - BCE or MSE - and the metric - Accuracy or MAE - of the training set should always be close to that of the validation set.

\par Fig. \ref{cv} shows the evolution of the loss function and the metric (in this case, the Mean Absolute Error) for 10 epochs, for the model trained above. If both curves of the validation set (dashed lines in Fig. \ref{cv}) are above the training set's (full lines), then there is underfitting, otherwise there is overfitting. The train and validation curves should ideally stay as close to each other as possible.

It is clear from Fig. \ref{cv} that there are some discrepancies between the full line (training set) and the dashed line (validation set). However, the variations do not represent a significant difference and it is therefore possible to validate this model and say that there is only slight overfitting, given the amount of noise thought to be present in the data.

\subsection{Comparison to Traditional Methods}

\begin{figure}[t]
\centering
\hspace*{-.5cm}  
\includegraphics[width=1.05\columnwidth]{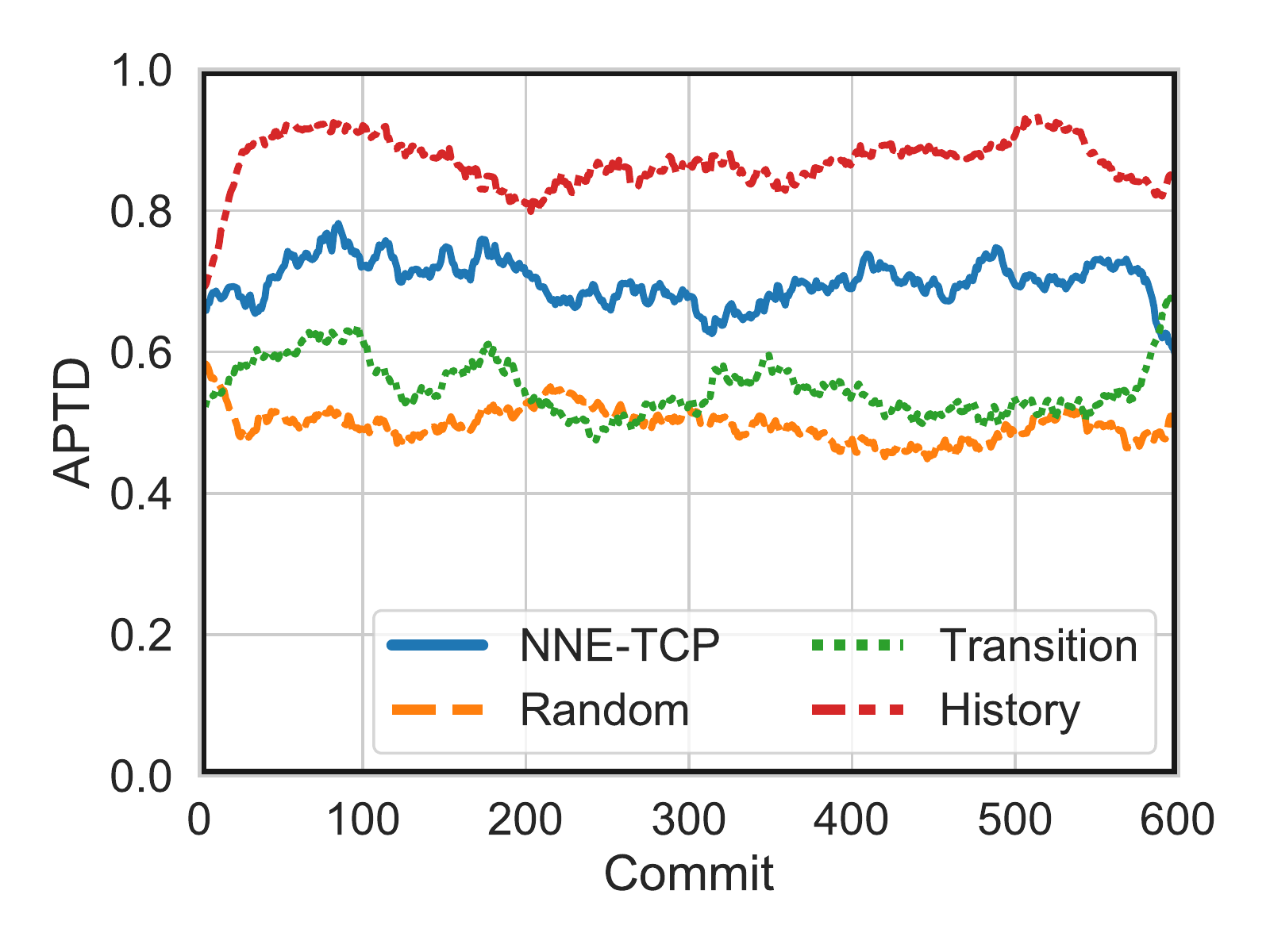}
\caption{APTD Trend of NNE-TCP and traditional methods for the test set. Each line is obtained by calculating the rolling average over a 50 commit window (for a total of 600 commits).}
\label{comparison}
\end{figure}

% Focus of this section
Having calibrated and trained the NNE-TCP model to produce the best APTD value, we are now able to study how this new approach performs when compared to the three traditional methods described earlier. The results of this comparison are depicted in Fig. \ref{comparison}.

%Line Plot

Out of the four methods presented, \textit{History} is the one that shows higher APTD trends, followed by our approach NNE-TCP and, further down, \textit{Transition} and \textit{Random}, respectively. Table \ref{tab} shows the average value for the APTD and the root-mean-squared error (RMSE) - the average of the quadratic difference between two curves - between NNE-TCP and the other traditional methods.

\begin{table}[b]
\centering
\normalsize
\begin{tabular}{lcc}
\hline
\textbf{Method}     & \multicolumn{1}{r}{\textbf{Mean APTD}} & \textbf{RMSE} \\ \hline
\textit{NNE-TCP}    & 0.70                                   & 0             \\
\textit{Random}     & 0.50                                   & 0.21          \\
\textit{Transition} & 0.59                                   & 0.16          \\
\textit{History}    & 0.87                                   & 0.17          \\ \hline
\end{tabular}
\caption{Performance comparison between TCP methods. The RMSE value is calculated in relation to NNE-TCP}
\label{tab}
\end{table}

\par As expected for a baseline method, \textit{Random} (yellow line) has an approximately constant APTD trend of around $0.5$, meaning that, on average, relevant test cases are not ranked in the beginning nor the end of the test schedule, but are rather uniformly distributed.
\par Looking at the \textit{Transition} method (green line) it is possible to see a slight improvement relative to \textit{Random}. Although this method only considers the frequency of past transitions for each test case it is able to assign less relevance to stable tests, i.e. test that almost do not cause transitions, have low priority and are therefore ranked at the bottom of the schedule. However, due to the large amount of tests, this method lacks the resources to make more meaningful and targeted schedules.

\par The \textit{History} method (red line) was able to achieve the highest trend of the four methods, with the assumption that tests that suffered transitions recently, are more likely to transition again. This assumption, also present in \textit{Transition}, allows for stable tests to be ranked at the bottom of the test schedule, but in a differentiating manner. For example, in \textit{Transition} if two test cases only transition once, but one of them did so very recently, the latter will be more relevant. Hence, stable test are weighted down by the time elapsed since their last execution. In short, the longer a test remains stable, the less relevant it becomes and, consequently the lower priority it has. Additionally, when a regression is detected, it is expected that it will be fixed soon, since the bug-source has clearly been identified, causing a progression. For this reason, a test that transitioned recently through a regression is more likely to transition again through a progression. All these factors help explain the success of the \textit{History} method.

% limitations
Nevertheless, when regressions occur, the longer a test has been failing, the larger its progression probability becomes, since more time has passed for developers to pin-point and fix the faulty commit. \textit{History} is unable to encapsulate this effect and is also limited in detecting newly introduced regressions, since it does not take into account the available information of each commit (e.g. the modified files), but rather relying only on its heuristic to only prioritize tests that have transitioned recently.

%nne
\par As we turn our attention to NNE-TCP (blue line) it is evident that it presents a significant improvement relative to the \textit{Random} and \textit{Transition} prioritization methods, by providing a targeted and commit-focus approach, that directly investigates which files were modified and, from the learned mapping to tests, is able to cherry-pick the tests with the most affinity.

However, it should be said that \textit{History} shows better prioritization capability than NNE-TCP, by only taking into account a test case's history of transitions. Notwithstanding, we have reasons to believe that NNE-TCP has not reached its full potential. Ideally, with a more abundant amount of quality data, a crystal clear relation between modified files and test cases could be found, with minimized uncertainty. With this information, when a commit is made, the algorithm will be fed with the files modified and know exactly which tests are more likely to affected, eliminating unnecessary executions. This way, the problem of not being able to detect newly introduced regressions would be settled, enabling quick transition detections, leading to early fixes. 

\par It is worth noticing that NNE-TCP is a novel approach that is evaluated by the equally novel APTD metric and, to the best of our knowledge, there are no other ML frameworks whose performance can be compared to NNE-TCP. Thus, it is of the utmost importance that we validate the model against other traditional methods, that do not learn from experience.

\par In conclusion, the NNE-TCP approach performs better than \textit{Transition} and \textit{Random}, and worse than \textit{History}. Nonetheless, we strongly believe that there are enough evidences to consider NNE-TCP as viable alternative to address the Test Case Prioritization problem and to be implemented in an industrial environment.

\subsection{Entity Representation}
\begin{figure*}[t]
    \centering
    \begin{minipage}[b]{0.45\linewidth}
            \centering    
            \hspace*{-1.5cm}
            \includegraphics[width=1.25\columnwidth]{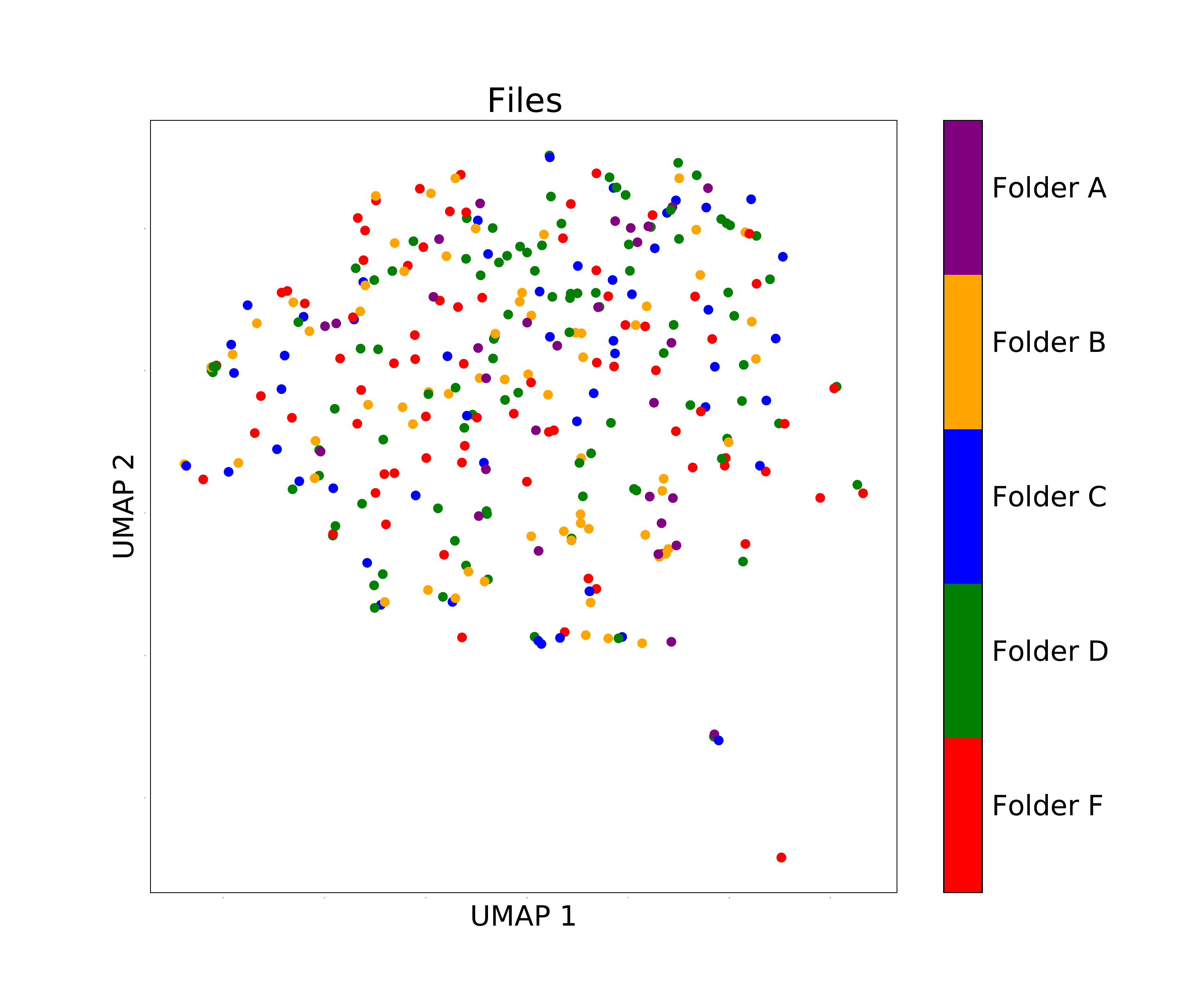}
        \end{minipage}
        \quad
        \begin{minipage}[b]{0.45\linewidth}
            \centering   
            \hspace*{-0.5cm} \includegraphics[width=1.25\columnwidth]{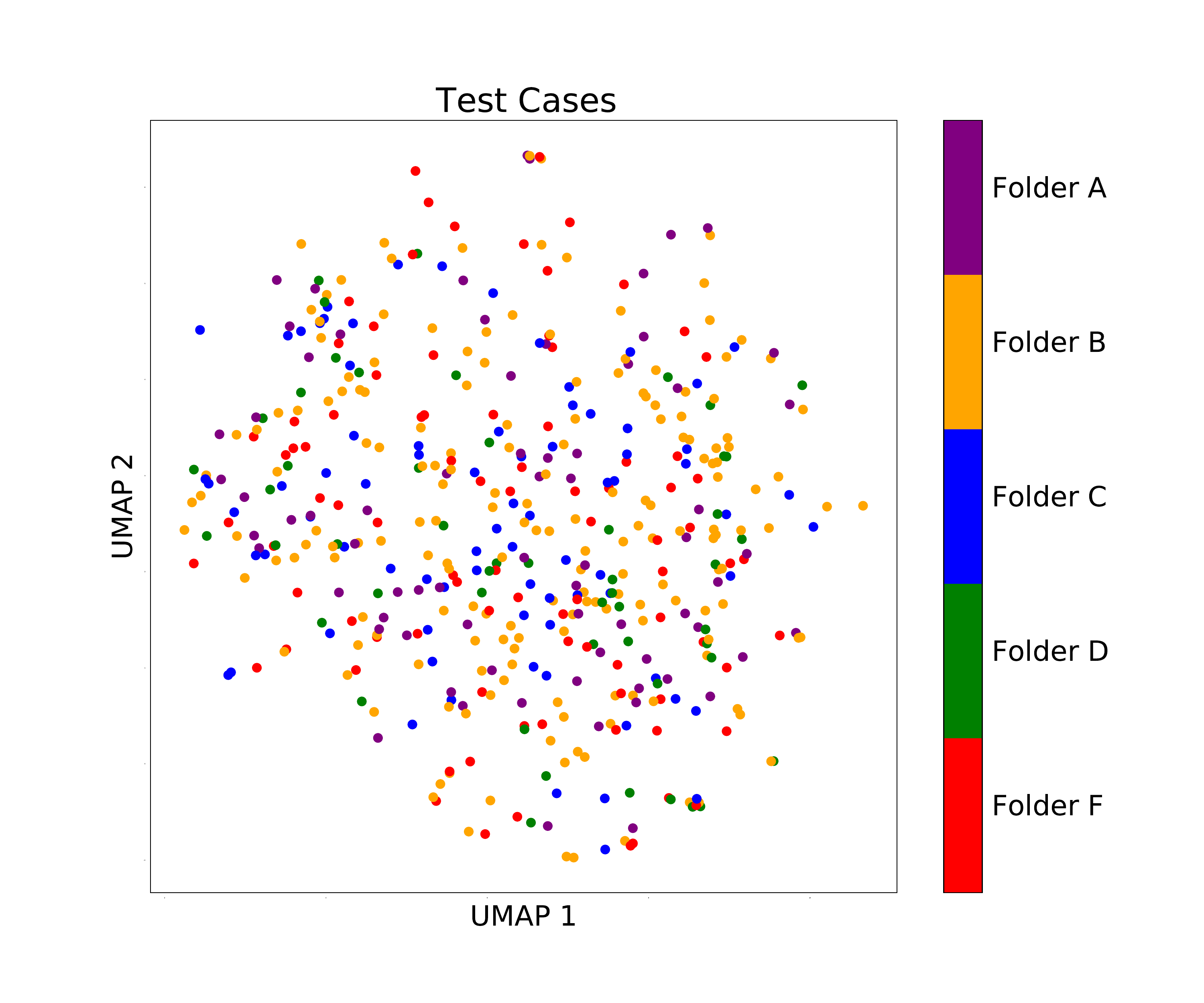}
        \end{minipage}
    \caption{Labelled Embeddings with UMAP technique. Labels correspond to the five most populated folders where files/test cases are stored in the system}
    \label{umap}    
\end{figure*}
% UMAP description
One of the advantages of using embeddings is that, after solving a Supervised Learning problem, the results can be represented in low dimensional space with UMAP \cite{umap}.
UMAP aims to find a low dimensional projection of the embeddings, while maintaining the same high-dimensional topological structure. This allows us to project multi-dimensional entities into two dimensions, granting us the possibility to observe some clusters of files and test cases with the same behaviour.

%Figure Description
Fig. \ref{umap} shows the embedding representation for files and test cases, obtained by using UMAP \cite{umap} and by labelling files and tests according to the corresponding folder where they are stored in the system.
In two dimensional space, UMAP has two components, which correspond to the plot's axes in Fig. \ref{umap}. Each dot represents a file or a test case. Dots that are close to each other represent files/test cases with similar behaviour.

With these results, we can see that in the test case projection, some dots overlap. This is a clear indicator of test case redundancy, i.e. test cases that suffer transitions concurrently, meaning they must have a very high degree of similarity. This information can then be used by a test engineer to inspect these particular test cases and, if applicable, clean the redundant ones. Removing redundancies is a crucial step to avoid wasting resources unnecessarily and saving time in regression testing.
\par Additionally, each dot has a corresponding label, helping us see if the current organization in folders of files and test cases is correlated with the behaviour of transitions. In Fig. \ref{umap}, only five folders - the ones with most frequent files and tests - were  considered from the set of all folders, to facilitate visualization. Because we can't observe any single color clusters in the projections, it is possible to conclude from the plot that there is no correlation between the folder where files/tests are stored and whether they cause/suffer transitions together. It should be noted that there were cases where files and tests' folder names were corrupted or unavailable.
\par Furthermore, there is the possibility to apply a Clustering algorithm to the obtained embeddings, where each cluster contains elements of similar behaviour, unlocking an array of possible improvements for TCP. By selecting one test from each cluster, we could possibly achieve maximum code coverage, i.e. covering most amount of source code with a particular subset of test cases, in the shortest amount of time or consuming the least amount of resources.

Finally, one very important aspect of entity representation is knowing which files cause transitions together. In software development, oftentimes there is a massive interdependence between files, e.g. external libraries, large projects, etc., and, sometimes, these are the source of many introduced regressions \cite{Uber}. These interdependencies are paper-thin and burdensome to find manually. From the embedding results, if two or more file dots overlap in Fig. \ref{umap}, then there is a strong indicator that the files have a relation between them.

\subsection{Threats to Validity}
\textbf{Internal.} The first threat to validity is associated with randomness when training the model, which was mitigated by using cross-validation, that trains the model multiple times. Furthermore, the dataset is relatively small, facing the number of test cases and modified files it encompasses. Collecting more data is a crucial step for Machine Learning models to learn better and more complex relations between inputs and outputs.
Another threat can derive from errors associated with our code implementation. Both \textit{Scikit-Learn} \cite{scikit-learn} and \textit{Keras} \cite{chollet2015keras} are well established frameworks used for Machine Learning. However, our implementation might contain inconsistencies and errors, which could lead to erroneous results. The code used in this work is available online at \textit{https://github.com/jlousada315/NNE-TCP} for validation purposes. 
\par \textbf{External.} This work is based on a single development environment, which is a major limitation considering the amount of real-world scenarios where CI is applied. This threat has to be addressed by running more experiments on data from several industries, with different development paces, team size, resources, etc.
As stated previously, the presence of noise in the dataset is a major setback that interferes with the algorithm's learning process. Human factors are one such example, and these can be mitigated by changing workplace practices, such as promoting a higher commit frequency and modifying fewer files on each commit rather than accumulating changes for long periods of time and then committing all at once. 

% To contribute to more data availability, publication of the dataset used in this paper is being authorized. 
\textbf{Construct.} In real-world complex CI systems, sometimes test cases change status due to extraneous reasons: system dependencies on several platforms can affect the outcome of a test, the infrastructure where tests are executed can suffer critical faults or malfunctions, some tests can fail to produce the same result each time they are executed (i.e. flaky tests). Therefore it is not certain that whenever a test case changed status it was due to a certain file being modified. Regarding the features used for training, our model only looks at the link between modified files and tests, whereas in real life, additional features may have a positive impact on the model's ability to make better predictions such as the commit author, execution history of each test, test age, etc. 
%To better validate the results achieved by NNE-TCP, it should be compared with other Machine Learning based Test Case Prioritization techniques described in the literature.

\subsection{Future Work}

The results of this work were the first step towards applying Embeddings in the context of TCP, by using file-test links as features, establishing a solid baseline for further studies.
\par Due to limited time and computer power, parameter tuning analysis was limited, but it can be further refined by exploring more combinations of parameters and measuring their impact on the APTD metric. Furthermore, our approach should take into consideration additional features, such as the commit author, to allow better understanding of expected transitions, shortening even more the delay between a software change and the information regarding whether it impacts the system's stability.
\par In terms of entity representation, embedding projection has the potential to provide a valuable insight about the system's structure, giving the chance to reorganize tests in different folders, grouped by similarity. Moreover, a clustering algorithm can be applied to group test cases. Test cases in the same cluster usually fail in similar situations, and therefore applying a subset of each cluster, rather than every test case, would avoid redundancies, without compromising code coverage, quickly grasping the status of the whole system.
\par Finally, NNE-TCP should be validated and tested against other datasets, so that it can become more flexible and adaptable to different contexts.

\section{Conclusion}

In this work we presented a novel approach to Test Case Prioritization using Machine Learning, called NNE-TCP. It combines Neural Network Embeddings with file-test links obtained from historical data, which, to the best of our knowledge, was done here for the first time. It is a lightweight and modular framework that not only predicts meaningful prioritizations, but also makes useful entity representations in the embedding space, grouping together similar elements, which is a valuable feature not usually available in other frameworks.

\par Our results point to a major improvement over some traditional TCP methods, which we called \textit{Random} and \textit{Transition}. However, NNE-TCP was unable to match the performance of a third, more complex traditional method, called \textit{History}. Nevertheless, we strongly believe that NNE-TCP has enough potential to reach higher performance levels. If a mapping between files and tests can be effectively learned by a data-driven approach, then only relevant tests will be executed, reducing feedback time. To validate this hypothesis, further experiments must be conducted on richer and cleaner datasets.
\par Finally, the ability to visualise embeddings in 2D space represents a valuable improvement over other commonly used methods, providing insights on the structure of the data. We showed that there does not seem to be exist any correlation between the folders where files/tests are stored and the similarity between the files/tests themselves. Notwithstanding, it is already possible to detect possibly redundant tests and discover dependencies between files.

                                  % on the last page of the document manually. It shortens
                                  % the textheight of the last page by a suitable amount.
                                  % This command does not take effect until the next page
                                  % so it should come on the page before the last. Make
                                  % sure that you do not shorten the textheight too much.

%%%%%%%%%%%%%%%%%%%%%%%%%%%%%%%%%%%%%%%%%%%%%%%%%%%%%%%%%%%%%%%%%%%%%%%%%%%%%%%%

%%%%%%%%%%%%%%%%%%%%%%%%%%%%%%%%%%%%%%%%%%%%%%%%%%%%%%%%%%%%%%%%%%%%%%%%%%%%%%%%

%%%%%%%%%%%%%%%%%%%%%%%%%%%%%%%%%%%%%%%%%%%%%%%%%%%%%%%%%%%%%%%%%%%%%%%%%%%%%%%%
\bibliographystyle{unsrt}
\bibliography{bibliography.bib}

\begin{thebibliography}{10}

\bibitem{santolucito2018statically}
Mark Santolucito, Jialu Zhang, Ennan Zhai, and Ruzica Piskac.
\newblock Statically verifying continuous integration configurations, 2018.

\bibitem{ShinThesis}
Yoo Shin.
\newblock {\em Extending the Boundaries in Regression Testing: Complexity,
  Latency, and Expertise}.
\newblock PhD thesis, King’s College London, 2009.

\bibitem{Ziftci}
Celal Ziftci and Jim Reardon.
\newblock Who broke the build? automatically identifying changes that induce
  test failures in continuous integration at google scale.
\newblock In {\em Proceedings of the 39th International Conference on Software
  Engineering: Software Engineering in Practice Track}, ICSE-SEIP ’17, page
  113–122. IEEE Press, 2017.

\bibitem{palma}
Francis Palma, Tamer Abdou, Ayse Bener, John Maidens, and Stella Liu.
\newblock An improvement to test case failure prediction in the context of test
  case prioritization.
\newblock pages 80--89, 10 2018.

\bibitem{Uber}
Sundaram Ananthanarayanan, Masoud~Saeida Ardekani, Denis Haenikel, Balaji
  Varadarajan, Simon Soriano, Dhaval Patel, and Ali-Reza Adl-Tabatabai.
\newblock Keeping master green at scale.
\newblock In {\em Proceedings of the Fourteenth EuroSys Conference 2019}, pages
  29:1--29:15. ACM, 2019.

\bibitem{yangJIT}
X.~{Yang}, D.~{Lo}, X.~{Xia}, Y.~{Zhang}, and J.~{Sun}.
\newblock Deep learning for just-in-time defect prediction.
\newblock In {\em 2015 IEEE International Conference on Software Quality,
  Reliability and Security}, pages 17--26, 2015.

\bibitem{learningfortcp}
Benjamin Busjaeger and Tao Xie.
\newblock Learning for test prioritization: An industrial case study.
\newblock In {\em Proceedings of the 2016 24th ACM SIGSOFT International
  Symposium on Foundations of Software Engineering}, FSE 2016, page 975–980,
  New York, NY, USA, 2016. Association for Computing Machinery.

\bibitem{liang}
Jingjing Liang, Sebastian Elbaum, and Gregg Rothermel.
\newblock Redefining prioritization: Continuous prioritization for continuous
  integration.
\newblock In {\em Proceedings of the 40th International Conference on Software
  Engineering}, ICSE ’18, page 688–698, New York, NY, USA, 2018.
  Association for Computing Machinery.

\bibitem{durelli}
V.~H.~S. {Durelli}, R.~S. {Durelli}, S.~S. {Borges}, A.~T. {Endo}, M.~M.
  {Eler}, D.~R.~C. {Dias}, and M.~P. {Guimarães}.
\newblock Machine learning applied to software testing: A systematic mapping
  study.
\newblock {\em IEEE Transactions on Reliability}, 68(3):1189--1212, 2019.

\bibitem{litrevtcp}
Heleno de~S.~Campos~Junior, Marco Ant\^{o}nio~P. Ara\'{u}jo, Jos\'{e} Maria~N.
  David, Regina Braga, Fernanda Campos, and Victor Str\"{o}ele.
\newblock Test case prioritization: A systematic review and mapping of the
  literature.
\newblock In {\em Proceedings of the 31st Brazilian Symposium on Software
  Engineering}, SBES'17, page 34–43, New York, NY, USA, 2017. Association for
  Computing Machinery.

\bibitem{7PrinciplesSoftTest}
B.~{Meyer}.
\newblock Seven principles of software testing.
\newblock {\em Computer}, pages 99--101, 2008.
\newblock Introduction on the fundamental pilars of software testing in a
  general way, useful in defining concepts.

\bibitem{APFD}
Gregg Rothermel, Roland~J. Untch, and Chengyun Chu.
\newblock Prioritizing test cases for regression testing.
\newblock {\em IEEE Trans. Softw. Eng.}, 2001.

\bibitem{Shinyoo}
Shin Yoo, Mark Harman, and Shmuel Ur.
\newblock Measuring and improving latency to avoid test suite wear out.
\newblock {\em IEEE International Conference on Software Testing, Verification,
  and Validation Workshops, ICSTW 2009}, 2009.

\bibitem{chen}
S.~{Chen}, Z.~{Chen}, Z.~{Zhao}, B.~{Xu}, and Y.~{Feng}.
\newblock Using semi-supervised clustering to improve regression test selection
  techniques.
\newblock In {\em 2011 Fourth IEEE International Conference on Software
  Testing, Verification and Validation}, pages 1--10, 2011.

\bibitem{kameiJIT}
Yasutaka Kamei, Emad Shihab, Bram Adams, Ahmed~E. Hassan, Audris Mockus, Anand
  Sinha, and Naoyasu Ubayashi.
\newblock A large-scale empirical study of just-in-time quality assurance.
\newblock {\em Software Engineering, IEEE Transactions on}, 39:757--773, 06
  2013.

\bibitem{Spieker}
Helge Spieker, Arnaud Gotlieb, Dusica Marijan, and Morten Mossige.
\newblock Reinforcement learning for automatic test case prioritization and
  selection in continuous integration.
\newblock {\em Proceedings of the 26th ACM SIGSOFT International Symposium on
  Software Testing and Analysis - ISSTA 2017}, 2017.

\bibitem{time-window}
Zhaolin Wu, Yang Yang, Zheng Li, and Ruilian Zhao.
\newblock A time window based reinforcement learning reward for test case
  prioritization in continuous integration.
\newblock In {\em Proceedings of the 11th Asia-Pacific Symposium on
  Internetware}, Internetware ’19, New York, NY, USA, 2019. Association for
  Computing Machinery.

\bibitem{chollet2015keras}
Francois Chollet et~al.
\newblock Keras, 2015.

\bibitem{osinga2018deep}
D.~Osinga.
\newblock {\em Deep Learning Cookbook: Practical Recipes to Get Started
  Quickly}.
\newblock O'Reilly Media, 2018.

\bibitem{umap}
Leland McInnes, John Healy, and James Melville.
\newblock Umap: Uniform manifold approximation and projection for dimension
  reduction, 2018.

\bibitem{scikit-learn}
F.~Pedregosa, G.~Varoquaux, A.~Gramfort, V.~Michel, B.~Thirion, O.~Grisel,
  M.~Blondel, P.~Prettenhofer, R.~Weiss, V.~Dubourg, J.~Vanderplas, A.~Passos,
  D.~Cournapeau, M.~Brucher, M.~Perrot, and E.~Duchesnay.
\newblock Scikit-learn: Machine learning in {P}ython.
\newblock {\em Journal of Machine Learning Research}, 12:2825--2830, 2011.

\end{thebibliography}

\end{document}